\documentclass[12pt]{article}
\usepackage{graphicx}
\usepackage[utf8]{inputenc} 
\usepackage[T1]{fontenc}
\usepackage{amsfonts,amssymb,amsthm,amsmath}
\usepackage[french,english]{babel}
\usepackage[lofdepth, lotdepth]{subfig}
				\usepackage[bbgreekl]{mathbbol}
				\usepackage[np]{numprint}
				\usepackage{mathcomp}
					
\begin{document}

\title{Quantum Theory and Special Relativity: when are physical properties attributed jointly to the system and to the context in which it is studied?}

\author{J. M. Vigoureux\\
Institut UTINAM, UMR CNRS 6213,\\Université de
Bourgogne Franche-Comté, 25030 Besançon Cedex,France.\\
{\em jean-marie.vigoureux@univ-fcomte.fr}}
\maketitle \fontsize{12}{24}\selectfont

\begin{abstract}
Alexia Auffèves and Philippe Grangier \cite{Grangier} proposed to modify the quantum ontology by requiring that physical properties are attributed jointly to the system and to the context in which it is embedded.
Comparing the formal structures of Quantum Theory and of Special Relativity may suggest a basis for such an interpretation of quantum theory.
\end{abstract}


\section{Introduction}
Attempts has been made to make quantum mechanics fully compatible with physical realism defined as the statement that "the goal of physics is to study entities of the natural world, existing independently from any particular observer's perception, and obeying universal and intelligible rules." Recently Alexia Auffèves and Philippe Grangier \cite{Grangier} proposed to modify the quantum ontology by requiring that physical properties are attributed \textit{jointly} to the system, and to the context in which it is embedded. \\
We underline that such a situation in which physical properties are attributed jointly to the system and to its context already exists in Physics and we think that comparing the formal structure of Special Relativity and Quantum Theory may shed a new light on some counter-intuitive features of quantum mechanics and could open a way to have quantum formalism and physical realism both correct and compatible.\\
In Special Relativity, some physical properties of a given system as measured by an observer both depend of the system and of the observer. Exactly as in the classical "perspective effect" we are used to in the everyday life and in the graphic arts, they should not any more be considered as properties of the system itself, but have to be jointly attributed to the system and to the context in which it is observed.\\
A first interesting feature of such a comparison is that, in Special Relativity, the resultant velocity which appears in its mathematical formulation as a "mathematical intrication" of the other velocities may appear \textit{as an usual addition of independant quantities when using hyperbolic geometry}. \\ 
Another one is that using hyperbolic geometry makes
any resultant velocity smaller than that of light (that is \textit{smaller than that which would be expected in classical mechanics}), exactly as using the spherical geometry (for quantum theory) makes the correlation fonction to appear greater than expected.\\

\subsection{Lorentz-Poincaré's and Einstein's : two interpretations of length contraction and time dilation}
There has been two different interpretations of relativistic time dilation and length contraction. The first one, that of Fitzgerald, Lorentz and Poincaré, considered that there is an actual physical contraction of a moving object originated from the action of an electromagnetic molecular force exerted by the ether. To this end, Poincaré introduced a sort of pressure (then called the "Poincaré stress") which were thought of as giving them a \textit{dynamical} explanation. According to this interpretation, moving objects are \textit{really} contracted by "ether vortices" and their clocks \textit{really} tick at a slower rate. This implies that there is no reciprocity between a moving observer and an observer at rest.
In that interpretation, relativistic effects being due to forces exerted by the ether, an object moving in ether and an object at rest in it are in fact in two different situations : one is moving through the ether (and it will consequently be shortened); the other is at rest in the ether and he consequently must keep its size.\\
\textit{This interpretation is clearly realist in that the observed effects are considered to exist independently from any particular observer's perception and that physical properties are totally attributed to the system.}\\
\indent In Einstein interpretation,
Lorentz transformations can be understood as a rotation in the $4$-dimensional Minkowski space. Exactly as a 3D rotation, a 4D rotation will not induce strain inside an object. Relativistic effects  can thus be illustrated as perspective effects as those we are used to in current life or in graphic arts \cite{perpective}. We know perpective due to the distance ; there also exists a perspective due to velocity. Exactly as an object appears to be smaller and smaller as its distance from the observer increases (a contextual property), it also seems smaller (and its time appears to be dilated) as its velocity with respect to the observer increases. The faster the relative velocity, the greater the magnitude of time dilation and of length contraction. We are not used to this "velocity-perspective" because in everyday life velocities are very small exactly as we would be not used to "distance-perspective" if we could see no further than two or three meters \cite{horizon}.\\
In perspective, a distant object is not \textit{really} smaller but it appears to be so because it is \textit{far} (context) \textit{from the observer}. The effect is both a \textit{property of the object itself} (it has its own "proper size"), and \textit{a property of my relation to it} (it is far from us).\\
In the same way, a moving object is not \textit{really} shortened, but it appears to be so when it is \textit{moving quickly with respect to the observer} (context). This effect is both a property of the object itself and a property of my \textit{relation} (relativity = relation) to it: it is moving with respect to me. \\
So, the system I am studying does exist with is proper characteristics but what I measure also depends on the context in which it is observed. Such a situation of course is compatible with realism.\\
\textit{So, in Einstein's interpretation, length contraction and time dilation are attributed jointly to the system and to the context in which it is "observed"} \cite{note reciprocite}.\\
To underline all this, let us quote Einstein who, 
on may 1911, wrote a paper in the Physikalische Zeitschrift on precisely this problem because he thought that using the words "time dilation" and "length contraction" could cause confusion. He wrote:
"\textit{the question of whether the Lorentz contraction exists or does not exist in reality is misleading, because it does not exist "in reality" insofar as it does not exist for an observer moving with the object. However, it does exist "in reality" in the sense that it could be detected by physical means by a non-comoving observer."} \cite{Einstein}.\\
The comparison between Lorentz-Poincaré's and Einstein's interpretations of Lorentz equations is interesting to understand our present point of view. Our aim in this paper is to show that such a comparison could help us to foresee a realist interpretation of quantum theory in which physical properties are attributed jointly to the system and to the context in which it is embedded.

 \section{Special Relativity}
Let us briefly recall some results of Special Relativity.
 Let us consider three moving galilean observers $O$, $O_1$ and $O_2$ and note $v_1$ the norm of the velocity of $O_1$ with respect to $O$, $v_2$ that of $O_2$ with respect to $O_1$ and $v_3$ that of $O_2$ with respect to $O$. As is well known, Lorentz transformations lead to the following equations (taking c = 1 and noting $\theta_1$ the angle between the two velocities $\vec{v}_1$ and $\vec{v}_2$)
 \begin{equation}\label{1}
\gamma_3 = \gamma_1 \gamma_2\,(1+ v_{1} \,v_{2}\, \cos{\theta_1})
\end{equation}
\begin{equation}\label{2}
\gamma_3 \,v_{3}\, \cos{\theta_3}= \gamma_1\,\gamma_2 \,(v_{1}+v_{2} \,cos{\theta_1})
\end{equation}
and
\begin{equation}\label{3}
v_{3}\, cos{\theta_3}= \frac{v_{1} + v_{2} \cos{\theta_1}}{1 + v_{1}\, v_{2} \cos \theta_1}
\end{equation}
where  $\gamma_i=1/ \sqrt{1- v_i^2/c^2}$ is the Lorentz factor.\\
These three equations can easily be found from a generalization in the complex plane of the usual composition law for parallel velocities  \cite{Vigoureux01}. Noting $v_i=\tanh\,a_i$ ($ v_i$ is the norm $\|\vec{v_i}\|$ of the velocity  and $a_i$ is the rapidity)  and  $\theta_i$ the orientation (in the plane of $\vec{v}_1$ and $\vec{v}_2$) of $\vec{v}_i$ with respect to an arbitrary x-axis of the reference frame of an observer, the composition law of velocity can be written
\begin{equation}\label{loiplus}
V_3=V_1\oplus V_2 = \frac{V_1 + V_2}{1 + \overline{V_1} V_2}.
\end{equation}
\noindent where each velocity $V_i$ is expressed by its polar form (velocities are expressed by small letters whereas polar forms are expressed by big ones)
\begin{equation}\label{defvitesse}
V_i=\tanh\displaystyle\frac{a_i}{2}\,e^{i\theta_i}
\end{equation}
and where the overbar means the complex conjugate. Mathematical properties of (\ref{loiplus}) have been given in \cite{Vigoureux93b}. All results resulting from the relativistic composition of  $\vec{v}_1$ and $\vec{v}_2$ can then be directly obtained from simple complex numbers operations by making use of eq.(\ref{loiplus}).\\
\indent Eq.(\ref{loiplus}) has two interesting consequences:\\
- first, it puts into evidence the \textit{non commutativity} of two non collinear boosts: the complex conjugate appearing in the denominator of (\ref{loiplus}) in fact shows that changing $V_1$ into $V_2$ in eq.(\ref{loiplus}) does not lead to the same resulting velocity. This non commutativity leads to the Thomas precession which is simply the commutator $[V_1,\,V_2]=V_1\oplus V_2-V_2 \oplus V_1$ \cite{Vigoureux93b} (we can note now that the same non commutativity appears with polarizers \cite{Lages08}: giving two polarizers $P_1$ and $P_2$, the final polarization of a light beam passing through $P_1$ and then through $P_2$ is obviously not the same as the polization obtained with $P_2$ and then $P_1$).\\
- second, the two velocities $V_1$ and $V_2$ in eq.(\ref{loiplus}), are in some way \textit{mathematically} entangled (as it is more commonly viewed we consider here entanglement as an algebraic concept) in the sense that it is not possible to mathematically separate the two velocities $V_1$ and $V_2$ in the resultant velocity $V_3$, the three referential frames so appearing as an inseparable whole and  the dynamic state of each referential frame so being \textbf{globally} described although they are separated and independant. The fact that it is not possible to separate the two velocities in the \textit{mathematical} expression of $V_3$ does not imply that $O_1$ and $O_2$ are \textit{physically} entangled. So it is not beacause two entities are not separable in a mathematic equation that they are so in the physical world. This will be useful in what follows.\\
As is well known, the three equations (\ref{1}, \ref{2}, \ref{3}) can be written by using hyperbolic geometry. Using the rapidities $a_i$ ($ v_i= \tanh a_i$), noting now $\theta$ the angle of $v_2$ with respect to $v_1$ and $\theta_3$ the angle of $v_3$ with respect to $v_1$, the three above equations in fact become (for the sake of clarity in what follows we now consider that the $v_1$ and $v_2$ are  the norm of the velocity of $O_1$ and $O_2$ with respect to $O$ and that $v_3$ is the norm of the velocity of $O_3$ with respect to $O_2$)
 \begin{equation}\label{hyperbolic1}
\cosh a_3= \cosh a_1  \cosh a_2 + \sinh a_1 \sinh a_2 \cos{\theta}
\end{equation}
\begin{equation}\label{hyperbolic2}
\sinh{a_3} \cos{\theta_3}= \sinh a_1  \cosh a_2+\cosh a_1 \sinh a_2 \cos{\theta}
\end{equation}
\begin{equation}\label{hyperbolic3}
\tanh{a_3} \cos{\theta_3}= \frac{\tanh{a_1}+ \tanh{a_2} \,\cos{\theta}}{1+\tanh{a_1} \, \tanh{a_2}\, \cos{\theta}}
\end{equation}
These three relations are basic equations of hyperbolic geometry. They are equivalent in that each one may lead to each others.
\begin{figure}[!h]
			\centering
			  \includegraphics[scale=0.5]{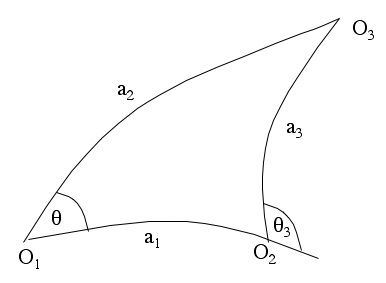}
			\caption{hyperbolic representation of the composition of velocities in Special Relativity: in the hyperbolic space, the resulting velocity is no more expressed in a "mathematically entangled" way (as in eq.\ref{loiplus})
but it appears as a simple addition of \textit{independant} quantities. The composed velocity is in fact obtained by placing the corresponding \textit{rapidities} head to tail and drawing $a_3$ from the free tail to the free head.
			\label{fig:1}}
		\end{figure}\\
It is interesting to note that the \textit{"mathematical entanglement"} (which appears in the above equations) dissapears in the hyperbolic space (see fig.1). In fact, when using hyperbolic geometry, the composition of velocities becomes a simple addition of entities having magnitudes $a_i$ (the rapidities) and directions $\theta_i$: for two velocities $v_1$ and $v_2$, the sum is in fact obtained by placing the corresponding rapidities $a_2$ and $a_3$ head to tail and drawing the vector from the free tail to the free head (fig.1) \cite{note} 

 So, it is possible to calculate the resulting velocity from two methods:\\
- one can use, for example,  eq.(\ref{loiplus}) where velocities appear to be \textit{mathematically} entangled, or\\
- we can use the usual addition law (vectorial sum) of \textit{independant rapidities} provide we use hyperbolic space.\\
Another important consequence of the hyperbolic structure of the velocity space is that adding velocities on the hyperbolic space makes their resultant to be naturally smaller than that expected in usual mechanics:\textbf{ $v\, <\, c$} (in fact $v= c \tanh a < c$). As we shall see, comparing the formal structures of quantum theory and of special relativity will show that using spherical geometry may \textit{naturally} lead in some case to correlations greater than that expected. \\
Let us add that knowing the velocity of $O_2$ with respect ot $O_1$ (eq.\ref{hyperbolic1}) requires that the velocities of $O_2$ and $O_1$ are defined \textit{in the same referential frame $O$} exactly as in quantum theory where EPR correlations can only be found if photons are
« twin photons » or if they have a common past.

\section{Quantum Theory}
 Consider two quantum systems A and B, with respective Hilbert spaces $H_A$ and $H_B$. The Hilbert space of the composite system is the tensor product $H_ A  \otimes H_B$.
Fixing a basis $\{|x\rangle_A, |y\rangle_A \}$ for $H_A$ and a basis $\{|x\rangle_B, |y\rangle_B\}$ for $H_B$ \textit{one} maximally entangled pure states can be 
  \begin{equation}\label{psi}
|\psi_{AB}> = a\, |x,x> + b\, |x,y>+c\, |y,x> + d\, |y,y>
\end{equation}
with the normalization relation
$$ |a|^2 +|b|^2+|c|^2+|d|^2 =1$$
Among them, let us consider the following one:
\begin{equation}\label{entangled}
|\psi> =\frac{1}{\sqrt{2}}(|x,x> + |y,y> \exp(j \beta))
\end{equation}
Noting $E(\theta_1,\, \theta_2) =  \cos 2\theta_3$ for the sake of homogeneity, this state leads to the correlation function
\begin{equation}\label{E}
E(\theta_1,\, \theta_2) =  \cos 2\theta_3= \cos 2\theta_1  \cos 2\theta_2+\sin 2\theta_1  \sin 2\theta_2 \cos \beta
\end{equation}
where $\theta_i$ define the orientations of polarizers. 
Noting $\theta=\theta_2-\theta_1$, the mean value of this correlation function
$$E(\theta)= \frac{1}{2} (\cos \beta+1) \cos 2 \theta $$
leads to EPR correlations when $\beta=0$.\\
Eq.(\ref{E}) is a well known relation in \textit{spherical trigonometry}. Exactly as in eqs.(\ref{hyperbolic1}, \ref{hyperbolic2}, \ref{hyperbolic3}), we could write the two other equivalent relations giving $ \sin 2\theta_3$ and $\tan 2\theta_3$.
However only eq.(\ref{E}) being used in quantum theory, it does not appear necessary to write these other equations here.\\
Eq.(\ref{E}) expresses the existing relationship between the three sides of a spherical triangle
  (fig.2) in which, as above, only appears a simple addition of \textit{independant} entities having magnitudes $2 \theta_i$ and directions $\beta_i$: the sum being obtained by placing the corresponding quantities head to tail and drawing $2 \theta_3$ from the free tail to the free head.\\
\begin{figure}[hbtp]
			\centering
			  \includegraphics[scale=0.3]{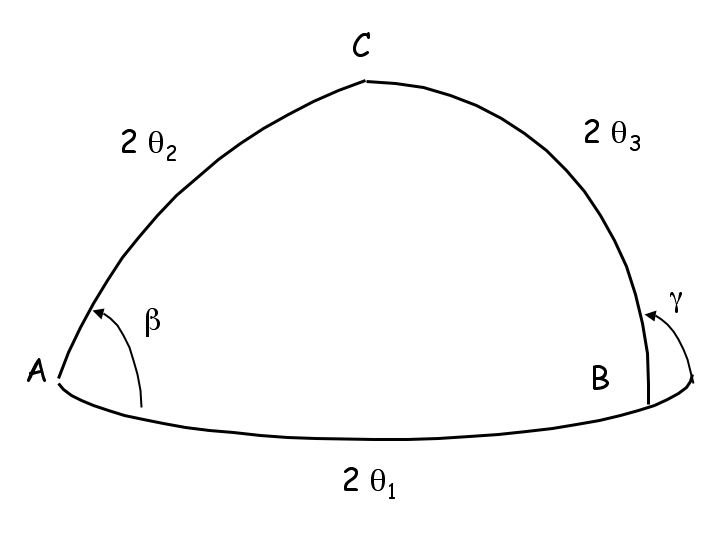}
			\caption{représentation of a correlation experiment. The EPR case corresponds to $\beta=0$ that is to the case where $\theta_3=\theta_2-\theta_1$.}
			\label{fig:2}
		\end{figure}
So, as in Special Relativity, it is also possible to calculate the correlation function from two methods:\\
- one can use eq.(\ref{E}) where quantities appear to be \textit{mathematically} entangled, or\\
- we can proceed as when adding vectors in euclidean space by adding \textit{independant} quantities in spherical space.\\
In eq.(\ref{E}) $\beta$ can be understood as the \textit{visibility} of intrication (but of course the intrication level remains the same whatever may be the value of $\beta$) \\
- when $\beta\,=\,0$ we find the EPR state.\\
- when $\beta\,=\,\pi/2$, although they are in an entangled state (eq.\ref{entangled}) the two photons do not produce such correlations (but they could with another experimental set-up).\\
These results can also be illustrated by taking an arbitrary reference axis $\theta_0$ for the polarization (fig.3).
\begin{figure}[hbtp]
			\centering
			  \includegraphics[scale=0.3]{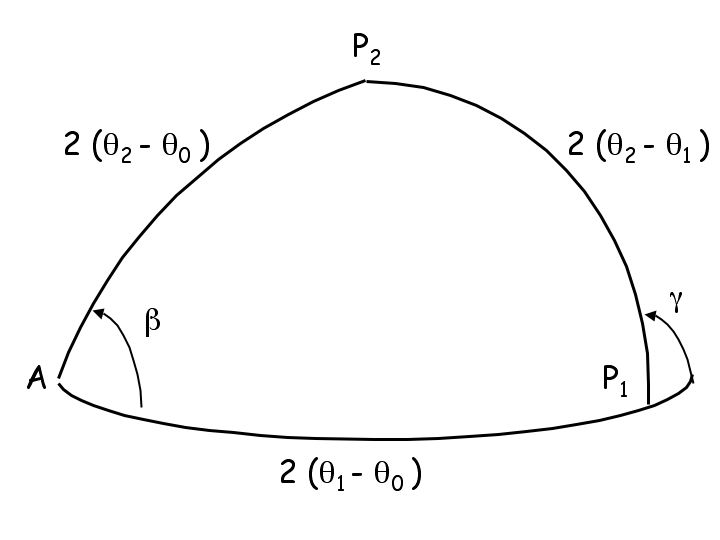}
			\caption{The case of entangled photons. The case of EPR experiment corresponds to $\beta=0$ ($\beta$ can be understood as the \textit{visibility} of intrication)}
			\label{fig:1}
\end{figure}

It can also be noted that using the spherical triangle shows that, in an EPR experiment, the fact that a given observer sees the axis of a first polarizer to be in the $\theta_1$ direction and the one of the second polarizer to be in the $\theta_2$ direction could not imply that \textit{for the photons}, the axis of the first polarizer with respect to the other is $\theta_2 - \theta_1$ (the fact that I measure $\theta_1$ for the polarisation of the first photon and $\theta_2$ for the other does not necessarily imply that one photon « sees » the polarization of the other to be in the direction $\theta_2 - \theta_1$; in a sherical triangle, the sum of the angles exceeds $\pi$ radians) exactly as in Special Relativity the fact that the velocity of a spaceship $S_1$ is $V_1$ in an inertial reference frame and that the velocity of another one $S_2$ is $V_2$ in the same frame does not imply that the velocity of $S_2$ with respect to $S_1$ is $V_2-V_1$

\section{conclusion}

Comparing the formal structures of Special Relativity and of Quantum Theory shows an example where physical properties can be attributed \textit{jointly} to the system, and to the context in which it is embedded.\\
We recognize that the present study deals with a too simple and very particular case and that a general study would be a lot more difficult. We however hope that this particular situation may shed a light on some counter-intuitive features of quantum mechanics and could open a way to understand how quantum formalism and physical realism may be both correct and compatible.\\
In special relativity, the velocities of two spaceships may appear to be mathematically entangled whereas their are independent (the real factual situation of a moving observer is independent of what happens with the other which is spatially separated from the former but their velocities, in the referential frame of an another observer are mathematically intricated).
 In the same way, quantum intrication could express in an inseparable way and for a given observer the behaviour of two independent particles. In correlation experiments, quantum theory may appear as a way of writing what a photon is "seeing" of the other exactly as, in special relativity, the composition law of velocities is only a mean to calculate in the referential frame of a given observer what a moving referential is "seeing" of another one.


\begin{thebibliography}{99}
\bibitem{horizon} To complete the parallel, it can also be noted that the velocity-perspective has its horizon line which is the velocity of light exactly as the distance-perspective has its horizon line which represents objects "infinitely" far away.
\bibitem{note} However, note that in the hyperbolic space of velocities, the parallelogram rule (which is only valid in euclidean space) is no more valid: as expected, the composition of two velocities depend on the order in which they are considered; here also appears the Thomas-Wigner angle.
\end{thebibliography}
\end{document}